{}
{}

\documentclass[prd,reprint,superscriptaddress]{revtex4-1}

\newif\ifpublic\publictrue

\usepackage{fancyhdr}
\ifpublic\else
\fi
\newif\ifworking\workingtrue

\usepackage{mathtools,mathrsfs,amsbsy,amssymb,latexsym,amsfonts,amscd,
amsmath}
\usepackage{bm}
\usepackage{graphicx}
\usepackage[usenames,dvipsnames]{color}
\usepackage[normalem]{ulem}
\usepackage{natbib}
\usepackage{xcolor}
\definecolor{linkcolor}{rgb}{0,0,0.6}
\usepackage[ pdftex,colorlinks=true,
pdfstartview=FitV,
linkcolor= linkcolor,
citecolor= linkcolor,
urlcolor= linkcolor,
hyperindex=true,
hyperfigures=false]
{hyperref}

\def\showkeysrefformat#1{{\normalfont\tiny\ttfamily#1}}
\makeatletter
\def\SK@@ref#1>#2\SK@{%
{\@inlabelfalse\leavevmode\vbox to\z@{%
\vss\SK@refcolor\rlap{\vrule\raise .75em%
\hbox{\showkeysrefformat{#2}}}}}}
\makeatother

\allowdisplaybreaks[3]

\begin{document}

\title{Kaluza-Klein Spectrometry for Supergravity}

\newcommand{\be}{\begin{equation}}
\newcommand{\ee}{\end{equation}}
\newcommand{\ben}{\begin{displaymath}}
\newcommand{\een}{\end{displaymath}}
\newcommand{\bea}{\begin{eqnarray}}
\newcommand{\eea}{\end{eqnarray}}
\newcommand{\nn}{\nonumber}
\newcommand{\non}{\nonumber\\}
\newcommand{\bean}{\begin{eqnarray*}}
\newcommand{\eean}{\end{eqnarray*}}
\newcommand{\beqs}{\begin{eqnarray}}
\newcommand{\eeqs}{\end{eqnarray}}

\author{Emanuel Malek}
\email{emanuel.malek@aei.mpg.de}
\affiliation{Max-Planck-Institut f\"{u}r Gravitationsphysik (Albert-Einstein-Institut), 
   	Am M\"{u}hlenberg 1, 14476 Potsdam, Germany}

\author{Henning Samtleben}
\email{henning.samtleben@ens-lyon.fr}
\affiliation{Univ Lyon, Ens de Lyon, Univ Claude Bernard, CNRS,
    Laboratoire de Physique, F-69342 Lyon, France}

\begin{titlepage}

\end{titlepage}

\begin{abstract}
Exceptional field theories yield duality covariant formulations of supergravity. We show that they provide a highly efficient tool to compute the Kaluza-Klein mass spectra associated to compactifications around various background geometries relevant for string theory and holographic applications. This includes geometries with little to no remaining symmetries, hardly accessible to standard methods. As an illustration, we determine the masses of some higher Kaluza-Klein multiplets around warped geometries corresponding to some prominent ${\cal N} = 2$ supersymmetric AdS vacua in maximal supergravity. 

\end{abstract}

\maketitle

\setcounter{equation}{0}

An old and central problem in theories with extra dimensions is the determination of the mass spectrum of higher-dimensional fluctuations around a given compactification background. From a phenomenological point of view, this is central to the question of which particles are observable in lower dimensions. In particular, massless scalar fields in general contradict particle physics observations, while scalar fields of negative mass squared indicate a vacuum instability which jeopardizes the entire compactification scenario. This is particularly relevant for vacua whose stability is not controlled by supersymmetry arguments. In the holographic context, the full Kaluza-Klein spectrum around particular geometries with anti-de Sitter (AdS) factors carries vital information about the conformal dimensions of gauge invariant operators in the dual gauge theory.

Computation of Kaluza-Klein mass spectra, in general, is a highly nontrivial problem, which requires linearization and diagonalization of the higher-dimensional field equations expanded in terms of the eigenmodes of suitable Laplacian operators on the internal manifold. There are only particular scenarios where this problem has been fully solved. For manifolds with large isometry group and preserving major fractions of supersymmetry, fluctuations organize themselves into (semi-)short multiplets of the superalgebra of background isometries, such that the internal harmonics are controlled by group theory and the mass eigenvalues are essentially determined by the residual quantum numbers. This has backed the early work on the Kaluza-Klein spectra of the maximally supersymmetric backgrounds AdS$_4\times S^7$~\cite{Englert:1983rn,Sezgin:1983ik,Biran:1983iy,Casher:1984ym}, and AdS$_5\times S^5$ \cite{Gunaydin:1984fk,Kim:1985ez}. Examples of compactifications with less supersymmetry yet controlled by the coset structure of the internal spaces include \cite{Ceresole:1999zs,Fre:1999gok}. 

For general manifolds, the problem is far more complicated. The mass spectrum of the spin-2 sector shows some universal pattern and can be determined from a wave equation depending only on the background geometry, not on the supergravity matter fields \cite{Bachas:2011xa}. This has been further exploited in \cite{Passias:2016fkm,Pang:2017omp,Gutperle:2018wuk,Dimmitt:2019qla}. By contrast, the fluctuation equations for the lower-spin fields, notably the scalar fluctuations, generically depend on the non-metric details of the background solution, such as the non-vanishing background fluxes of $p$-forms. Moreover, these fluctuations mix together the various matter fields such that mass eigenstates have to be meticulously disentangled. This renders the general analysis highly non-trivial. 

On supersymmetric backgrounds, the information from the spin-2 sector may be extrapolated to some of the other matter fields upon exploiting the multiplet structure of the fluctuations \cite{Klebanov:2008vq,Klebanov:2009kp,Gutperle:2018wuk}. However, this approach offers only partial access to the Kaluza-Klein spectrum as it remains restricted within the spin-2 multiplets. Moreover, even for those mass eigenstates sitting inside spin-2 multiplets the laborious task of identifying the corresponding fluctuations within the higher-dimensional theory remains. This is indispensable for any holographic application.

In this letter, we will show that exceptional field theory \cite{Hohm:2013pua} offers a very powerful tool to solve this problem for large classes of examples. Exceptional field theories yield a duality covariant formulation of higher-dimensional supergravity theories. They have proven instrumental in constructing consistent truncations from higher-dimensional supergravities \cite{Hohm:2014qga,Malek:2017njj}. In particular, they offer a constructive way to obtain the non-linear reduction Ans\"{a}tze of the higher-dimensional theory in terms of the fields of a lower-dimensional gauged supergravity, such that all solutions of the lower-dimensional theory induce solutions of the higher-dimensional field equations.
In this letter, we will demonstrate that this construction may be naturally extended to also produce the form of the higher-dimensional fluctuations around any solution uplifted from the lower-dimensional theory.
In particular, we find that the formalism automatically disentangles the higher-dimensional fluctuation equations which allows us to obtain compact and universal formulas for the mass matrices of the infinite Kaluza-Klein towers.
We briefly illustrate the formalism for a couple of prominent ${\cal N}=2$ supersymmetric AdS vacua in maximal supergravity. 
Details will appear in an upcoming paper \cite{MalekSamtleben}.

As an example, we discuss the domain wall solution of $D=5$ gauged supergravity that uplifts to a solution of type IIB supergravity interpolating between the maximally supersymmetric AdS$_5\times S^5$ and a warped  AdS$_5\times M^5$ geometry~\cite{Freedman:1999gp}.  The latter background is the conjectured holographic dual of the infrared (IR) fixed point of the renormalization group (RG) flow triggered by a mass deformation of ${\cal N} = 4$ super-Yang-Mills theory. The internal manifold $M^5$ is a deformation of the round sphere $S^5$, preserving only U(2) isometries and breaking supersymmetry down to one quarter. Moreover all $p$-forms in ten dimensions acquire non-vanishing background fluxes. Accordingly, this background is not amenable to standard techniques of harmonic analysis. We show that within exceptional field theory, the full Kaluza-Klein spectrum around this background can be computed with the mass eigenstates neatly expressed in terms of the harmonics of the round $S^5$,
and we give the explicit results for the first level.
The same pattern applies to many other holographic backgrounds including \cite{DHoker:2007zhm,Klebanov:2009kp,Gutperle:2018wuk,Fischbacher:2010ec,Bobev:2013yra,Godazgar:2014eza,Guarino:2015jca,Bobev:2019dik}
(to name a few)
and will allow us to extract their hitherto unknown Kaluza-Klein spectra.

Let us start by briefly reviewing the structure of the relevant E$_{6(6)}$ exceptional field theory (ExFT); for details we refer to \cite{Hohm:2013pua,Baguet:2015xha}. 
This ExFT is a universal formulation of all higher-dimensional supergravities in terms of the fields of $D=5$
maximal supergravity. Its bosonic sector 
\bea
\left\{g_{\mu\nu}, {\cal M}_{MN},{\cal A}_\mu{}^M, {\cal B}_{\mu\nu\,M} \right\}
\;,\quad&&
\mu=0, \dots,4\;,
\nonumber\\
&&{}
M=1, \dots, 27
\;,
\label{fields}
\eea
comprises an external and an internal metric $g_{\mu\nu}$, ${\cal M}_{MN}$, respectively, with the latter parametrizing 
the coset space E$_{6(6)}/{\rm USp}(8)$, together with vector and tensor fields, ${\cal A}_\mu{}^M$ and ${\cal B}_{\mu\nu\,M}$,
transforming in the
${\bf 27}$ and ${\bf 27'}$ of the group E$_{6(6)}$, respectively.
All fields formally live on a $(5+27)$-dimensional exceptional space-time 
with coordinates $\{x^\mu, Y^M\}$, subject to the section constraint
 \be
  d^{MNK}\,\partial_N \otimes \partial_K  \ = \ 0 \,, 
 \label{section_constraint}
 \ee  
with the symmetric E$_{6(6)}$ invariant $d$-symbol $d^{MNK}$.
The ExFT Lagrangian resembles the generic structure of maximal supergravity in five dimensions:
\bea
{\cal L}  &\equiv &   \widehat{R}+\frac{1}{24}\,g^{\mu\nu}{\cal D}_{\mu}{\cal M}^{MN}\,{\cal D}_{\nu}{\cal M}_{MN}\label{LE6}
\\
 &&{}-\frac{1}{4}\,{\cal M}_{MN}{\cal F}^{\mu\nu M}{\cal F}_{\mu\nu}{}^N
 +|g|^{-1/2}{\cal L}_{\rm top}-V({\cal M},g) \,.
 \nonumber
\eea
Here, derivatives are covariantized ${\cal D}_\mu = \partial_\mu - \mathbb{L}_{A_\mu}$ w.r.t.\ 
generalized diffeomorphisms acting as
\bea
\!\!\!\!\!\!\!\mathbb{L}_\Lambda {\cal M}_{MN} = 
\Lambda^K \partial_K {\cal M}_{MN}   
+ 12\, \partial_K \Lambda^L\,\mathbb{P}^K{\!}_L{}^P{\!}_{(M}{\cal M}_{N)P}
\,,
\label{deltaM}
\eea
with the projector onto the adjoint representation
\bea\label{projector}
\mathbb{P}^M{\!}_N{}^K{\!}_L
 &=&
\frac1{18}\,\delta_N^M\delta^K_L + \frac16\,\delta_N^K\delta^M_L
-\frac53\,d_{NLR}\,d^{MKR}
\,.\;\;
\label{Padj}
\eea
The non-abelian field strengths read
\bea
\!\!\!\!{\cal F}_{\mu\nu}{}^M \equiv
 2\,\partial_{[\mu} {\cal A}_{\nu]}{}^M 
-\big[{\cal A}_{\mu}, {\cal A}_{\nu}\big]^M_{\rm E}
+ 10\,  d^{MNK} \partial_K {\cal B}_{\mu\nu N}\,, \;\;\;
\label{defF}
\eea
with the non-abelian E-bracket $[,]_{\rm E}$ derived from the action (\ref{deltaM}),
and the coupling to 2-forms ${\cal B}_{\mu\nu M}$ required in order to
achieve gauge covariance. The topological term in (\ref{LE6}) is obtained by integrating
\bea
d{\cal L}_{\rm top} &\propto& 
d_{MNK}\,{\cal F}^M \wedge  {\cal F}^N \wedge  {\cal F}^K
\qquad
\nonumber\\
&&{}
-40\, d^{MNK}{\cal H}_M\,  \wedge \partial_N{\cal H}_K
\,,
\label{CSlike}
\eea
with ${\cal H}_M$ denoting the non-abelian 3-form field strength of the tensor fields ${\cal B}_M$.
Finally, the potential $V({\cal M},g)$ is a gauge invariant combination of
terms bilinear in internal derivatives acting on internal and external metric.

The section constraint (\ref{section_constraint}) implies that fields depend on no more than six of the 
internal coordinates. Upon inequivalent choices of the physical internal coordinates among the $\{Y^M\}$,
the Lagrangian (\ref{LE6}) reproduces that of the full $D=11$ and ten-dimensional IIB supergravity, respectively.
Specifically, the IIB coordinates are identified upon breaking E$_{6(6)}$ down to ${\rm GL}(5)\times {\rm SL}(2)$
with
\bea
\left\{Y^M\right\} &\longrightarrow&
\left\{
Y^m, Y_{kmn}, Y_{m}{}^{\alpha}, Y^\alpha
\right\}\;,\nonumber\\
&&{}
m=1, \dots, 5\,;\;\;
\alpha=1,2
\;,
\label{GL5}
\eea
and restricting all field dependence to $\{x^\mu, Y^m\}$.
Upon analogous decomposition of the ExFT fields (\ref{fields})
under ${\rm GL}(5)\times {\rm SL}(2)$, followed by proper
on-shell dualizations and field redefinitions, one can establish the precise
dictionary to recover the full field content of the ten-dimensional IIB theory \cite{Baguet:2015xha}.

ExFT has proven a powerful tool for the construction of consistent truncations. The reduction formulas
for the ExFT fields (\ref{fields}) take the form of a generalized Scherk-Schwarz Ansatz
  \bea
  \label{SSforms}
 g_{\mu\nu}(x,Y) &=& \rho^{-2}(Y)\,g_{\mu\nu}(x)\;,
 \nonumber\\
 {\cal M}_{MN}(x,Y) &=& U_{M}{}^{\underline{K}}(Y)\,U_{N}{}^{\underline{L}}(Y)\,M_{\underline{KL}}(x)\;, 
\nonumber\\
  {\cal A}_{\mu}{}^{M}(x,Y) &=& \rho^{-1}(Y)(U^{-1})_{\underline{N}}{}^{M}(Y)\, A_{\mu}{}^{\underline{N}}(x) \;, 
  \nonumber\\
  {\cal B}_{\mu\nu\,M}(x,Y) &=& 
  \,\rho^{-2}(Y)\, U_M{}^{\underline{N}}(Y)\,B_{\mu\nu\,\underline{N}}(x)
  \;,
 \eea
in terms of an E$_{6(6)}$ twist matrix $U$ and a weight factor~$\rho$. 
The consistency conditions for the twist matrix are most compactly expressed as conditions of
generalized Leibniz parallelizability \cite{Lee:2014mla}
  \begin{equation}
\mathbb{L}_{{\cal U}_{{\underline{M}} }} {\cal U}_{\underline{N}}  =
X_{\underline{MN}}{}^{\underline{K}} \, {\cal U}_{\underline{K}} 
\;,
\quad \mbox{for}\;\;
{\cal U}_{\underline{M}} \equiv \rho^{-1} U^{-1}_{\underline{M}}
\;,
\label{UUXU}
\end{equation}
with constant embedding tensor $X_{\underline{MN}}{}^{\underline{K}}$.
Once the twist matrix satisfies the consistency conditions (\ref{UUXU}), all dependence
on the internal coordinates factors out from the IIB equations of motion, which then reduce
to the equations of motion of maximal $D=5$ supergravity with gauging defined
by the embedding tensor (\ref{UUXU}).

The $S^5$ reduction of IIB supergravity is described by a particular twist matrix 
living in ${\rm SL}(6)\subset {\rm E}_{6(6)}$,
induced by the $6\times 6 $ matrix
\bea
\!\!\!\!(U^{-1})_A{}^{\hat{m}} &=&
\left\{
(U^{-1})_A{}^{0}  ,\,
(U^{-1})_A{}^{{m}}\right\} \nonumber\\
&=&
\mathring{\omega}^{1/3}
\left\{
\mathring\omega^{-1}\, {\cal Y}^A  ,\,
\mathring{g}^{mn}\partial_n {\cal Y}^A +4\,\mathring{\zeta}^m  {\cal Y}^A
\right\}
\,,\;\;\;\;
\label{U6}
\eea
in terms of  elementary sphere harmonics ${\cal Y}^A{\cal Y}^A=1$, 
($A=1, \dots, 6$), the round $S^5$ metric
$\mathring{g}_{mn} = \partial_m {\cal Y}^A \partial_n{\cal Y}^A$,
and the vector field $\mathring{\zeta}^n$ defined by
$\mathring\nabla_n\mathring{\zeta}^n = 1$.
 The weight factor is given by $\rho=\mathring{\omega}^{-1/3}$
 in terms of the metric determinant $\mathring{\omega}^2 = {\rm det}\, \mathring{g}_{mn}$\,.
With the embedding of ${\rm SL}(6)\times {\rm SL}(2)\subset {\rm E}_{6(6)}$
described by the breaking 
of the ${\bf 27}$ as
\bea
A^{\underline{M}} &\longrightarrow&
\left\{
{A}^{AB}, A_{A\alpha}
\right\}
=
\left\{
{A}^{[AB]}, A_{A\alpha}
\right\}
\;,
\label{break27}
\eea
the induced ${\rm E}_{6(6)}$ twist matrix $({\cal U}_{(S^5)})_{\underline{M}}{}^M$ satisfies (\ref{UUXU})
with the non-zero components of the embedding tensor given by 
\bea
X_{\underline{MN}}{}^{\underline{K}} &:& 
\left\{
\begin{array}{l}
X_{AB,CD}{}^{EF} =
 2\,\sqrt{2}\,\delta_{[A}{}^{[E}\delta_{B][C}\delta_{D]}{}^{F]}\;, 
\\[1ex]
 X_{AB}{}^{C\alpha}{}_{D\beta} =
-\sqrt{2}\,\delta_{[A}{}^{C}  \delta_{B]D}\,\delta^\alpha_\beta  \;.
 \end{array}
 \right.\;\;\;
 \label{XMNK}
\eea
In particular, the twist matrix satisfies the relation
\bea
{\cal U}_{\underline{M}}{}^{N} \partial_N &=&
{\cal K}_{\underline{M}}{}^m\,\partial_{m}
\;,
\label{UK}
\eea
with the ${\rm SO}(6)$ Killing vector fields 
\bea
{\cal K}_{AB}{}^m &=& \sqrt{2} \,\mathring{g}^{mn}{\cal Y}^{[A} \partial_n {\cal Y}^{B]}
\;,\nonumber\\
{\cal K}^{A\alpha\,m} &=& 0
\;.
\label{KillingSO6}
\eea
Within the reduction Ansatz (\ref{SSforms}), the AdS$_5\times S^5$ solution of IIB supergravity
takes the simple form
\bea
g_{\mu\nu}(x)=(g_{{\rm AdS}_5})_{\mu\nu}(x)\;,\quad
M_{\underline{MN}}(x)&=&\delta_{\underline{MN}}
\;,
\label{backExFT}
\eea
with vectors and tensors vanishing. Fluctuations of the IIB theory around this background solution
organize into an infinite tower of short Kaluza-Klein multiplets of increasing masses~\cite{Gunaydin:1984fk,Kim:1985ez}.
The Ansatz (\ref{SSforms}) describes the full non-linear embedding into ten dimensions of the lowest (massless) Kaluza-Klein multiplet
that carries the field content of $D=5$ maximal gauged supergravity \cite{Gunaydin:1985cu}, such that every solution of the D=5 theory
lifts to a solution of the IIB field equations \cite{Baguet:2015sma}.

In this letter, we address the higher Kaluza-Klein multiplets.
In the standard formulation of IIB supergravity,
fluctuations are formulated in terms of appropriate sphere harmonics. For example, a ten-dimensional
scalar field gives rise to a tower of $D=5$ scalar fields
\bea
\phi(x,Y) &=& \sum_\Sigma\,{\cal Y}^\Sigma\,\varphi_\Sigma(x)
\;,
\eea
accompanying the scalar harmonics ${\cal Y}^\Sigma$ on the round $S^5$,
i.e.\ the sphere functions on which the Killing vector fields (\ref{KillingSO6}) have a linear action
\bea
{\cal K}_{\underline{M}}{}^m \partial_m {\cal Y}^\Sigma &=&
-{\cal T}_{\underline{M}}{}^{\Sigma\Omega}\,{\cal Y}^\Omega
\;,
\label{KTY}
\eea
with ${\rm SO}(6)$ generators ${\cal T}_{\underline{M}}$. Specifically, for $S^5$ these harmonics
can be expressed as polynomials in the elementary harmonics ${\cal Y}^A$ as
\bea
\left\{{\cal Y}^\Sigma \right\} &=&
\left\{1, {\cal Y}^A,   {\cal Y}^{A_1A_2},  \dots, {\cal Y}^{A_1\dots A_n}, 
\dots \right\} 
\;,
\label{harmList}
\eea
where we denote by ${\cal Y}^{A_1\dots A_n}\equiv {\cal Y}^{((A_1} \dots {\cal Y}^{A_n))}$ 
traceless symmetrization.
The index $\Sigma$ thus runs over the tower of symmetric vector representations $[n,0,0]$ of
${\rm SO}(6)$\,.
For the fields of non-vanishing spin, the relevant harmonics on coset spaces such as $S^5={\rm SO}(6)/{\rm SO}(5)$
can be classified and determined by group theoretical methods~\cite{Salam:1981xd}.

The main result which we will exploit in this letter, is the observation that in terms of the ExFT variables (\ref{fields}),
fluctuations around the background (\ref{backExFT}) are most compactly expressed by combining the non-linear embedding
of the lowest multiplet (\ref{SSforms}) with the infinite tower of  scalar harmonics ${\cal Y}^\Sigma$.
More precisely, for vector and tensor fields, the full set of IIB fluctuations is described by the generalization of (\ref{SSforms}) to
\bea
  {\cal A}_{\mu}{}^{M} &=& \rho^{-1}(U^{-1})_{\underline{N}}{}^{M}\, 
  \sum_\Sigma \,{\cal Y}^\Sigma
  A_{\mu}{}^{\underline{N},\Sigma}(x) \;, 
  \nonumber\\
  {\cal B}_{\mu\nu\,M} &=& 
  \,\rho^{-2}\, U_M{}^{\underline{N}}\,
    \sum_\Sigma\, {\cal Y}^\Sigma
    B_{\mu\nu\,\underline{N},\Sigma}(x)
\;,
\label{fluc1}
\eea
with the sum running over scalar harmonics \eqref{harmList}. For the external and the internal metric,
the Ansatz 
\bea
g_{\mu\nu} &=& \rho^{-2} \left(
(g_{{\rm AdS}_5})_{\mu\nu}(x)
+  \sum_\Sigma {\cal Y}^\Sigma\,
  h_{\mu\nu,\Sigma}(x)  \right)
  \;,
  \nonumber\\
{\cal M}_{MN} &=& 
U_{M}{}^{\underline{K}}\,U_{N}{}^{\underline{L}}\,
\left(
\delta_{\underline{KL}}
+ \sum_\Sigma \,{\cal Y}^\Sigma\,j_{\underline{KL},\Sigma}(x)
\right)
\;,
\label{fluc2}
\eea
is given in terms of fluctuations further restricted by the fact that these metrics
parametrize the coset spaces ${\rm GL}(5)/{\rm SO}(5)$ and ${\rm E}_{6(6)}/{\rm USp}(8)$, respectively.

The conditions \eqref{UUXU} satisfied by the twist matrices
ensure that with this Ansatz to linear order in the fluctuations, all dependence on the internal coordinates still factors out
from the equations of motion. The latter thus
reduce to linear five-dimensional differential equations. In particular, in the IIB field equations, internal derivatives act throughout via the combination \eqref{UUXU} and \eqref{UK}, i.e.\ their action on the sphere harmonics is realized by the action (\ref{KTY}) of Killing vector fields. 
Consequently, the resulting equations do not mix fluctuations over different ${\rm SO}(6)$ representations $\Sigma$.
The same structure underlies the ExFT supersymmetry transformations~\cite{Musaev:2014lna}. 
As a result, all fluctuations in (\ref{fluc1}), (\ref{fluc2}) associated with a fixed
${\rm SO}(6)$ representations $\Sigma=[n,0,0]$ combine into a single 1/2-BPS multiplet BPS$[n]$. 
This is to be contrasted with the structure in the original IIB variables:
after evaluating the products of the sphere harmonics ${\cal Y}^\Sigma$ with the $Y$-dependent twist matrices in (\ref{fluc1}), (\ref{fluc2})
and translating the ExFT fields back into the IIB supergravity fields, fluctuations of the original IIB fields combine linear combinations 
of different mass eigenstates originating from different BPS multiplets.

E.g.\ with (\ref{fluc1}), the twist matrix (\ref{U6}), and harmonics (\ref{harmList}), one reads off the 
component ${\cal B}_{\mu\nu}{}^\alpha$ of the ten-dimensional two-form as
\bea
{\cal B}_{\mu\nu}{}^\alpha &=& 
\sum_{n=0}^\infty
{\cal Y}^{A}{\cal Y}^{C_1 \dots C_n}\,
B_{\mu\nu}{}^{A\alpha,C_1 \dots C_n}(x)
\nonumber\\
&=&
\sum_{n=0}^\infty \Big(
B_{\mu\nu}{}^{((C_1\alpha,C_2 \dots C_{n}))}
\nonumber\\
&&{}+
\frac{n+1}{2\,(n+3)} \, B_{\mu\nu}{}^{A\alpha,AC_1 \dots C_{n}}
 \Big)\;
 {\cal Y}^{C_1 \dots C_{n}}\,
\;,
\eea
mixing in its fluctuations different mass eigenstates originating from
multiplets BPS$[n-1]$ and BPS$[n+1]$,
in precise accordance with the result of \cite{Kim:1985ez}.

We now compute the mass matrices by plugging the Ansatz \eqref{fluc1}, \eqref{fluc2} into the equations of motion. For the tensor fields, the Lagrangian (\ref{LE6}) gives rise to the first order
duality equations
 \bea
d^{PML}\partial_L  \left({\cal M}_{MN} {\cal F}^{\mu\nu N}
 +\kappa\,  \epsilon^{\mu\nu\rho\sigma\tau}\,
  {\cal H}_{\rho\sigma\tau M}\right) \ = \ 0
\;,
 \label{dualityFH}
\eea
with $\kappa^2 \equiv \frac5{32}$\,. 
With the field strengths (\ref{defF}) carrying a St\"uckelberg type couplings to the two-forms,
linearization and gauge fixing of (\ref{dualityFH}), together with an evaluation of internal derivatives
on twist matrices and scalar harmonics, gives rise to fluctuation equations for topologically massive two forms
\bea
3\,\partial_{[\mu} B_{\nu\rho]\,\underline{M}\alpha} &=& \frac12\, \epsilon_{\mu\nu\rho\sigma\tau}\,
M^{\underline{M}\Sigma,\underline{N}\Omega}\,B^{\sigma\tau}{}_{\underline{N}\alpha}
\;,
\eea
with the antisymmetric mass matrix given by
\bea
\!\!\!\!\!\!\!M^{\underline{M}\Sigma,\underline{N}\Omega} &\propto&
2\,d^{\underline{MKL}}\,X_{\underline{KL}}{}^{\underline{N}}\,\delta^{\Sigma\Omega} 
- 10\,   d^{\underline{MNK}}\, {\cal T}_{\underline{K}}{}^{\Sigma\Omega}
\,.
\label{massB}
\eea
It exhibits a very intriguing form as a superposition of the mass matrix of the $D=5$ supergravity
describing the lowest Kaluza-Klein multiplet with the SO(6) action (\ref{KTY}) on the scalar harmonics.

A similar, although more lengthy computation, linearizing the second order vector field equations
descending from (\ref{LE6}) yields the vector mass operator
\bea
M_{\underline{M}\Sigma,\underline{N}\Omega}&\propto&
\frac13\,
X^{\rm s}_{\underline{ML}}{}^{\underline{K}}
\,   X^{\rm s}_{\underline{NK}}{}^{\underline{L}}
\,
\delta^{\Sigma\Omega} 
\nonumber\\
&&{}
+
2\left(
X^{\rm s}_{\underline{MK}}{}^{\underline{N}}
-   X^{\rm s}_{\underline{NM}}{}^{\underline{K}}
  \right) {\cal T}_{\underline{K},\Omega\Sigma} 
\nonumber\\
&&{}
-6\,
 \left(
 \mathbb{P}^{\underline{K}}{}_{\underline{M}}{}^{\underline{L}}{}_{\underline{N}}
 +\mathbb{P}^{\underline{M}}{}_{\underline{K}}{}^{\underline{L}}{}_{\underline{N}}
\right)
{\cal T}_{\underline{L},\Omega\Lambda} {\cal T}_{\underline{K},\Lambda\Sigma}  
\nonumber\\
&&{}
+\frac{8}{3}\,{\cal T}_{\underline{N},\Omega\Lambda} {\cal T}_{\underline{M},\Lambda\Sigma}  
\;,
\label{massA}
\eea
in terms of the symmetrized $X^{\rm s}_{\underline{MN}}{}^{\underline{K}}
\equiv
X_{\underline{MN}}{}^{\underline{K}}+X_{\underline{MK}}{}^{\underline{N}}$\,,
and the adjoint projector (\ref{Padj}).
Again, this formula combines the mass matrix of the $D=5$ supergravity with the SO(6) action (\ref{KTY}).
Finally, a similar formula can be derived for the scalar fluctuations (\ref{fluc2}). For the spin-2 fluctuations, the Ansatz \eqref{fluc2} yields the simple mass formula $M_{\Sigma,\Omega} \propto {\cal T}_{\underline{M},\Sigma\Lambda} {\cal T}_{\underline{M},\Lambda\Omega}$, coinciding with \cite{Bachas:2011xa,Dimmitt:2019qla}.

In order to diagonalize the mass matrices level by level in 
the harmonics (\ref{harmList}), we need to evaluate the formulas (\ref{massB}), (\ref{massA})
with the explicit form of the embedding tensor (\ref{XMNK}) as well as the explicit expressions
for the E$_{6(6)}$ tensor $d^{\underline{KMN}}$ and the SO(6) action (\ref{KTY}) in the basis (\ref{break27})
\bea
d^{AB,CD,EF} &=& \tfrac1{\sqrt{80}}\,\varepsilon^{ABCDEF}\;,
\nonumber\\
d^{AB}{}_{C\alpha,D\beta} &=& \tfrac1{\sqrt{5}}\, \delta^{AB}_{CD}\,\varepsilon_{\alpha\beta}
\;,
\nonumber\\
({\cal T}_{AB})_{CD} &=& {\sqrt{2}}\,\delta_{C[A}\delta_{B]D}
\;.
\eea
It is then a straightforward exercise to determine the mass eigenvalues of the different SO(6)
representations at level $\Sigma=[n,0,0]$, i.e.\ of the different irreps of
$A^{AB,C_1\dots C_n}$, $A_{A\alpha}{}^{C_1\dots C_n}$, etc. We summarize the result in Table~\ref{Tab:specS5n}, which agrees with \cite{Gunaydin:1984fk,Kim:1985ez}. It moreover confirms that the propagating fluctuations
described by the Ansatz (\ref{fluc1}), (\ref{fluc2}) at level $\Sigma=[n,0,0]$ precisely span the bosonic part of a single 1/2-BPS multiplet.
Moreover, the Ansatz together with the dictionary of ExFT into IIB supergravity allows us to directly localize the different
components of the BPS multiplets within the IIB theory.

\begin{table}[t]
\begin{tabular}{c|cc}
fluctuation & SO(6) & $m^2$
\nonumber\\
\hline
$A_\mu{}^{A((B,C_1C_2\dots C_n))}$ &
$[n,1,1]$ & 
$n(n+2)$
\nonumber\\
$A_\mu{}^{B[C_1,C_2]C_3\dots C_n B}$ &
$[n-2,1,1]$ & 
$(n+2)(n+4)$
\nonumber\\
$A_\mu{}^{\alpha [A,C_1]C_2\dots C_n}$ &
$[n-1,1,1]$ 
& $(n+1)(n+3)$
\nonumber\\
\hline
$B_{\mu\nu}{}^{\alpha ((A,C_1\dots C_n))}$ &
$[n+1,0,0]$
& $(n+1)^2$
\nonumber\\
$B_{\mu\nu}{}^{\alpha B,BC_2\dots C_n}$ &
$[n-1,0,0]$
& $(n+3)^2$
\nonumber\\
$B_{\mu\nu}{}^{[AB,C_1]C_2\dots C_n}$ &
$[n-1,0,2]$~+~c.c.
& $(n+2)^2$
\nonumber\\
\hline
\end{tabular}
\caption{Mass spectrum (\ref{massB}), (\ref{massA}), on round $S^5$ at level $n$.}
\label{Tab:specS5n}
\end{table}

While this allows for a very compact rederivation of the known results for $S^5$, the construction has vastly more 
far-reaching applications. Since the reduction Ansatz is exact to all orders in the lowest multiplet, 
we may export the formulas to derive the
Kaluza-Klein spectrum around any solution of the $D=5$ supergravity.

As an example, we consider the ${\cal N}=2$ supersymmetric stationary point in the $D=5$ scalar potential 
conjectured to be dual to the IR fixed point of the RG flow triggered by a mass deformation of maximal super Yang-Mills theory~\cite{Freedman:1999gp}.
It is represented by a twist matrix
\bea
{\cal U}_{\underline{M}}{}^M &=& {\cal V}_{\underline{M}}{}^{\underline{N}}\,({\cal U}_{S^5})_{\underline{N}}{}^M
\;,
\label{UVU}
\eea
where the constant E$_{6(6)}$ matrix ${\cal V}_{\underline{M}}{}^{\underline{N}}$ identifies the location of the stationary point
on the coset manifold E$_{6(6)}/{\rm SU}(8)$. Accordingly, the fluctuation Ansatz (\ref{fluc1}), (\ref{fluc2}) holds with the
new twist matrix (\ref{UVU}), and the mass formulas (\ref{massB}), (\ref{massA}) hold with embedding tensor 
$X_{\underline{MN}}{}^{\underline{K}}$ and SO(6) generator ${\cal T}_{\underline{M}}$ dressed by the
 matrix ${\cal V}_{\underline{M}}{}^{\underline{N}}$.
Without going into details, we give the resulting mass spectrum around this vacuum 
at level 1 organized into multiplets of ${\rm SU}(2)\times{\rm SU}(2,2|1)$ (in the notation of \cite{Freedman:1999gp})
\bea
{\bf 0} &:& 
D( 1+\tfrac12\sqrt{37} , 0,0 ; 1)_{\mathbb{C}} 
+D( 1+\tfrac12\sqrt{61} , 0,0 ; 1)_{\mathbb{C}} 
\nonumber\\
&&{}
+D_{\rm S}(\tfrac92, \tfrac12, \tfrac12;1)_{\mathbb{C}}
+2\,D_{\rm S}(\tfrac92, \tfrac12, 0;-1)_{\mathbb{C}}
\nonumber\\
&&{}
+D(\tfrac92, \tfrac12, 0;1)_{\mathbb{C}}
\nonumber \,,\\[0.5ex]
{\bf \tfrac12} &:&
D( 1+\tfrac14\sqrt{145} , \tfrac12,\tfrac12 ; \tfrac12)_{\mathbb{C}} 
+D( 1+\tfrac14\sqrt{193} , 0,0 ; \tfrac12)_{\mathbb{C}}   
\nonumber\\
&&{}
+D(\tfrac{15}{4}, \tfrac12, 0; \tfrac12)_{\mathbb{C}}
+D(\tfrac{17}{4}, \tfrac12, 0; -\tfrac12)_{\mathbb{C}}
\nonumber\\
&&{}
+D_{\rm S}( \tfrac{15}{4}, 0, 0; \tfrac52 )_{\mathbb{C}} 
+D_{\rm S}( \tfrac{17}{4}, 0, 0; \tfrac32 )_{\mathbb{C}} 
\nonumber \,,\\[0.5ex]
{\bf 1} &:&
2\,D(1+\sqrt{7}, 0,0;0)
+D(1+\sqrt{7}, \tfrac12, 0;0)_{\mathbb{C}}
\nonumber\\
&&{}
+D_{\rm S}( \tfrac{7}{2}, \tfrac12, 0; 1 )_{\mathbb{C}}
+D_{\rm S}( 3, \tfrac12, 0; 2 )_{\mathbb{C}}
\nonumber \,,\\[0.5ex]
{\bf \tfrac32} &:&
D_{\rm S}( \tfrac{9}{4},0, 0; \tfrac32 )_{\mathbb{C}}
\;.
\eea
It is a non-trivial consistency check that the masses obtained by our formulas consistently combine
into ${\rm SU}(2,2|1)$ multiplets.
By $D_{\rm S}$ we denote semi-short multiplets whose energy saturates one of the unitarity bounds~\cite{Flato:1983te}.
In contrast, the energy of the long multiplets cannot be deduced from representational arguments,
but only from direct computation as presented here. Complex multiplets $D(E_0,j_1,j_2;r)_{\mathbb{C}}$
come in pairs with their conjugates $D(E_0,j_2,j_1;-r)_{\mathbb{C}}$.
It is interesting to note that the complex multiplet 
$D_{\rm S}( \tfrac{7}{2}, \tfrac12, 0; 1 )_{\mathbb{C}}$ contains two massless scalars.

As a second example, we study the ${\rm U}(3)$ invariant ${\cal N}=2$ AdS$_4$ vacuum identified in the scalar potential of maximal
$D=4$ gauged supergravity \cite{deWit:1982bul,Warner:1983vz}, conjectured to be the holographic dual of certain
matter-coupled Chern-Simons theories. The general multiplet structure of the Kaluza-Klein spectrum has been
analyzed in \cite{Klebanov:2008vq} by group theoretical methods, which however do not give access to the masses of
the long multiplets. Adapting the above mass formula (\ref{massA}) to E$_{7(7)}$ ExFT~\cite{Hohm:2013uia} allows us to straightforwardly determine
the full spectrum. We list our result for the energies of the long OSp$(2|4) \times {\rm SU}(3)$  multiplets 
appearing at the first level (in the notation of \cite{Klebanov:2008vq})
\bea{}
 {\rm LVEC} : [0,0] &:& E_0=\tfrac12+\tfrac12\sqrt{33} \,,
\nonumber\\{}
 {\rm LGRAV} :[1,0]+[0,1] &:& E_0=\tfrac12+\tfrac16\sqrt{145} \,,
\nonumber\\{}
 {\rm LGINO}:[1,0]+[0,1] &:& E_0=\tfrac{17}{6} \,,
\nonumber\\{}
 {\rm LVEC}:[1,0]+[0,1] &:& E_0=\tfrac12+\tfrac16\sqrt{217} \,,
\nonumber\\{}
{\rm LVEC}:[2,0]+[0,2]   &:& E_0=\tfrac73 \,,
\nonumber\\{}
 {\rm LGINO} : [1,1] &:& E_0=\tfrac12+\sqrt{3}
\;.
\eea
This extends the result of \cite{Klebanov:2009kp} for the long graviton multiplet to
all the long multiplets at this level.
In particular, the energy values we find establish
that there is no multiplet shortening occurring for these multiplets.

Our mass formulas (\ref{massB}), (\ref{massA}), thus offer direct access to the full Kaluza-Klein spectra around these
squashed and stretched spheres, hardly accessible to standard methods. As another intriguing application, the Ansatz (\ref{fluc1}), (\ref{fluc2}),
being exact in the lowest Kaluza-Klein multiplet, will allow us to compute within ExFT the holographic 2-point correlation functions of arbitrary operators
throughout the renormalization group flows \cite{Freedman:1999gp,Corrado:2001nv} described as domain wall solutions of the
lower-dimensional supergravity.

Further applications include similar analysis for the AdS vacua identified and studied in
\cite{DHoker:2007zhm,Klebanov:2009kp,Gutperle:2018wuk,Fischbacher:2010ec,Bobev:2013yra,Godazgar:2014eza,Guarino:2015jca,Bobev:2019dik}.
More generally, it will be interesting to combine the presented technology with recent numerical advances in the search for such vacua~\cite{Comsa:2019rcz}.
Of special interest are the non-supersymmetric AdS vacua whose stability so far could only been addressed within the lowest Kaluza-Klein multiplet. The technology presented here gives direct access to their notoriously difficult stability analysis. This will be particularly interesting in the light of the recent conjectures by Ooguri and Vafa on the absence of such vacua \cite{Ooguri:2016pdq}.
Another interesting direction is the generalization of this framework to vacua within consistent truncations preserving smaller fractions of supersymmetry \cite{Malek:2017njj}, giving access to yet larger classes of relevant AdS vacua.

\noindent\textbf{Acknowledgements}:  
We would like to acknowledge the Mainz Institute for Theoretical Physics (MITP) of the Cluster of Excellence PRISMA+ (Project ID 39083149) for hospitality while this work was initiated. EM is supported by the ERC Advanced Grant ``Exceptional Quantum Gravity'' (Grant No. 740209).

\end{document}